\documentclass{article}

\usepackage{arxiv}
\usepackage{xcolor}
\usepackage[utf8]{inputenc} 
\usepackage[T1]{fontenc}    
\usepackage{hyperref}       
\usepackage{url}            
\usepackage{booktabs}       
\usepackage{amsfonts}       
\usepackage{nicefrac}       
\usepackage{microtype}      
\usepackage{lipsum}		
\usepackage{graphicx}
\usepackage{natbib}
\usepackage{doi}

\title{SAMM (Segment Any Medical Model): A 3D Slicer Integration to SAM}

\author{ {Yihao Liu}\thanks{Equal Contribution} \\
	Department of Computer Science\\
	Johns Hopkins University\\
	Baltimore, MD \\
	\texttt{yliu333@jhu.edu} \\
	\And
	{Jiaming Zhang*} \\
	Department of Computer Science\\
	Johns Hopkins University\\
	Baltimore, MD \\
	\texttt{jzhan282@jhu.edu} \\
        \And
	{Zhangcong She*} \\
	Department of Mechanical Engineering\\
	Johns Hopkins University\\
	Baltimore, MD \\
	\texttt{zshe1@jhu.edu} \\
  \And
	{Amir Kheradmand} \\
	Department of Neurology\\
	Johns Hopkins University\\
	Baltimore, MD \\
	\texttt{akherad1@jh.edu} \\
  \And
	{Mehran Armand} \\
	Department of Computer Science\\
	Johns Hopkins University\\
	Baltimore, MD \\
	\texttt{marmand2@jhu.edu} \\
}

\renewcommand{\headeright}{}
\renewcommand{\undertitle}{}
\renewcommand{\shorttitle}{SAMM (Segment Any Medical Model): A 3D Slicer Integration to SAM}

\hypersetup{
pdftitle={SAMM (Segment Any Medical Model): A 3D Slicer Integration to SAM},
pdfsubject={},
pdfauthor={Yihao Liu, Jiaming Zhang, Zhangcong She},
pdfkeywords={SAM, Medical Image, 3D Slicer},
}

\begin{document}
\maketitle

\begin{abstract}
The Segment Anything Model (SAM) is a new image segmentation tool trained with the largest available segmentation dataset. 
The model has demonstrated that, with prompts, it can create high-quality masks for general images. 
However, the performance of the model on medical images requires further validation. 
To assist with the development, assessment, and application of SAM on medical images, we introduce Segment Any Medical Model (SAMM), an extension of SAM on 3D Slicer - an image processing and visualization software extensively used by the medical imaging community. 
This open-source extension to 3D Slicer and its demonstrations are posted on GitHub (https://github.com/bingogome/samm). 
SAMM achieves 0.6-second latency of a complete cycle and can infer image masks in nearly real-time.
\end{abstract}


\section{Introduction}
The advent of foundation models has led to significant progress in image analysis with potential for future advancements.  
SAM \citep{kirillov2023segment} is a revolutionary foundation model for image segmentation and has already shown the capability of handling diverse segmentation tasks. 
SAM especially prevails in its generalization capability compared with the existing fine-tuned models that are trained on specific domains. 
Thus, SAM holds significant promise for application in medical image segmentation: The advantage lies in adapting it to address the inherent inter-subject variations and low signal-to-noise ratio commonly found in medical images.

Medical image segmentation is the task to separate different structures within an image. The segmentation results can then be used to detect the region of interest or reconstruct 3-dimensional anatomical models \citep{sinha2020multi}. The existing AI-based segmentation methods, however, do not fully bridge the domain gap among different anatomies and different imaging techniques \citep{wang2020deep}. 
This introduces challenges for training, as AI systems need to be trained on anatomy- and task-specific datasets.
Therefore, a universal tool would be valuable if it can be applied across all image modalities and various anatomical structures.
However, the universal tool must also overcome a series of critical challenges including (but not limited to) data privacy, ethics, expenses, scalability, data integrity, and validation \citep{gao2023}. 
In contrast, pretrained SAM can perform a new segmentation task without being fine-tuned on task-specific datasets \citep{kirillov2023segment}. 
This feature makes SAM a promising segmentation tool for various image modalities with less effort. 

Despite the extensive use of AI in medical imaging, the application of foundation models, such as SAM, remains largely unexplored.
The application of these models would require resolving differences in the structure between medical images and non-medical images. 
3D Slicer \cite{fedorov2012}, as a widely-used open-source software, provides off-the-shelf functionalities to manipulate 2D and 3D medical images, using consistent and user-friendly interfaces and visualization tools. 
Therefore, we introduce the Segment Any Medical Model (SAMM) that incorporates 3D Slicer and SAM, to assist the investigation of applying general foundation models to medical images.

\section{Methodology}
\subsection{Overall Architecture}

\begin{figure}
    \centering
    \includegraphics[width=0.9\textwidth]{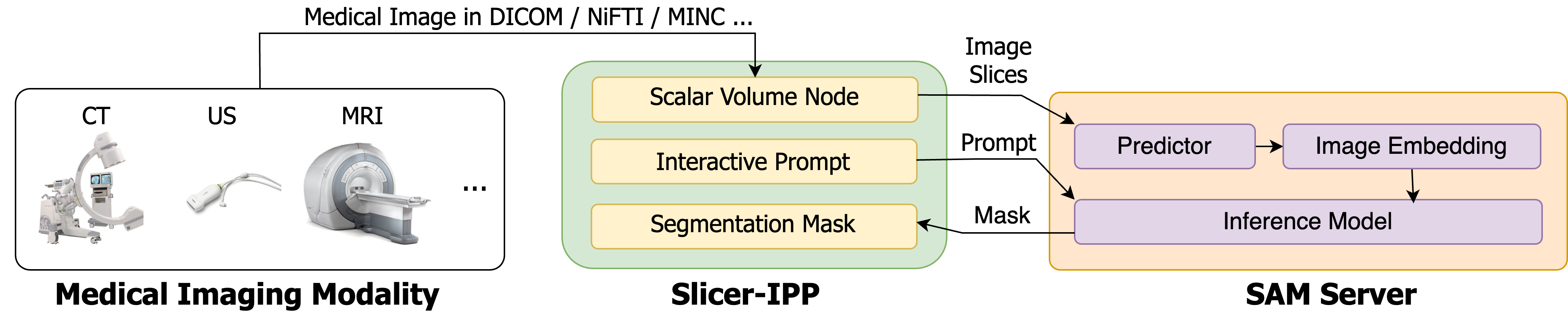}
    \caption{Overall architecture of the integration of 3D Slicer and SAM. Images from different modalities, such as computed tomography (CT), magnetic resonance imaging (MRI), or ultrasound (US), are contained in a scalar volume node. The node (vtkMRMLScalarVolumeNode) is a class in the Visualization Toolkit (VTK) that represents a volume of the scalar data within the Medical Reality Markup Language (MRML) framework.}
    \label{fig:overall}
\end{figure}

Figure \ref{fig:overall} presents the overall architecture of SAMM, which consists of a SAM Server and an interactive prompt plugin for 3D Slicer (Slicer-IPP). 
SAM Server first loads the pretrained SAM. It runs in parallel with Slicer-IPP and keeps monitoring the requests sent from 3D Slicer. On the other side, Slicer-IPP handles all the image slices with the built-in interfaces of 3D Slicer. Then, it processes all the slices and send them to SAM Server to compute their embeddings. Subsequently, the embeddings of the slices are stored in a format of binary files in Random Access Memory (RAM) for efficient retrieval at the inference stage.  
Once the embeddings of all slices are ready, the user may start the segmentation using prompts. These prompts are the fiducial points placed on the 2D slice to indicate adding or removing the region. 
The prompt points are transmitted to the prompt encoder of SAM, and the inference stage starts synchronously. Here the prompt transmission and the SAM inference are synchronized. In such way, the mask generation responds to user action in real-time. 

Note that the image encoders run once per volume, rather than per prompt, which allows the users to segment the same image multiple times with different prompts in real-time.
Given that the initialization of image embedding occurs in advance, the subsequent mask generation process can be performed with small latency (Section \ref{sec:result}). 

The Slicer-IPP handles the transformations from the volume coordinates to 2D slice numbers and pixel coordinates. It can work out with discrepancies between the RAS (right, anterior, superior), IJK (slice ID in different views), and 2D image pixel coordinate systems by providing proper conversion functionalities. 
For instance, at an inference request, Slicer-IPP converts the coordinates of RAS to IJK to identify the image IDs. Slice-IPP then transmits the IDs along with the prompts to the SAM Server. With the coordinates converted, the masks of a 2D slice can be generated by SAM Server in the inference step. The coordinates of the generated masks are transformed back from the pixel to the RAS system, and are sent to Slicer-IPP for the visualization of the segmentation results.

\begin{figure}
    \centering
    \includegraphics[width=0.6\textwidth]{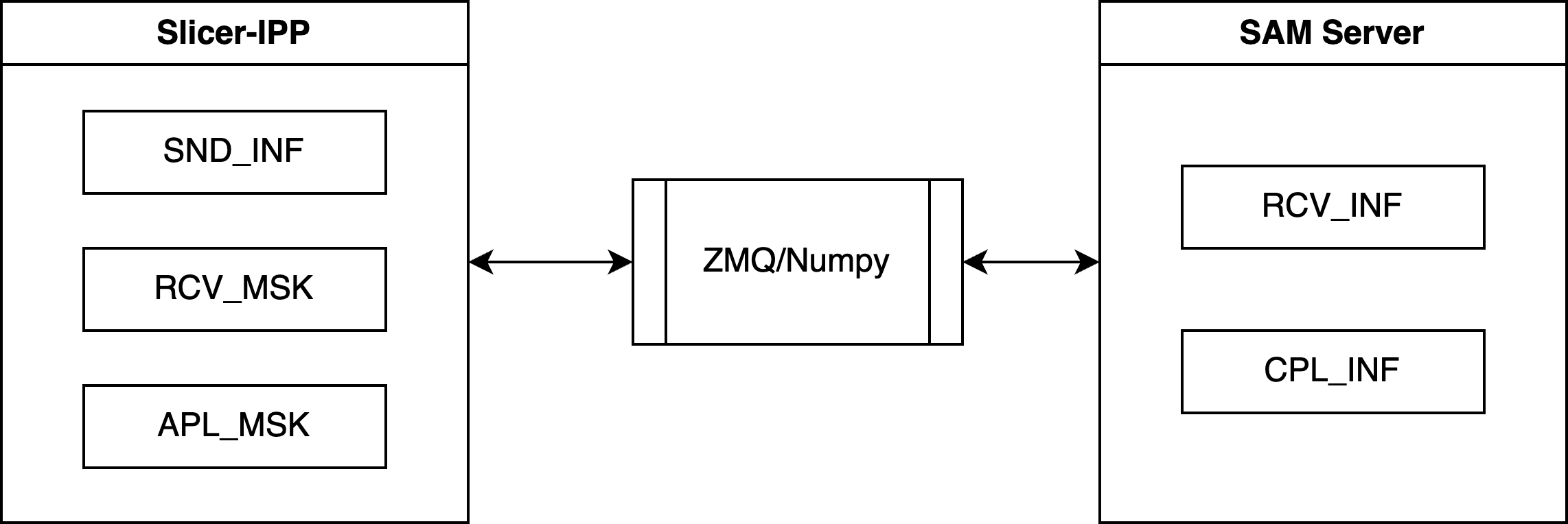}
    \caption{
    The affiliation of five tasks, namely are ``send inference request'' (SND\textunderscore{ }INF), ``receive inference request'' (RCV\textunderscore{ }INF), ``complete SAM inference'' (CPL\textunderscore{ }INF), ``receive mask transmission'' (RCV\textunderscore{ }MSK), and ``apply mask'' (APL\textunderscore{ }MSK). Slicer-IPP and SAM Server use ZMQ/Numpy memory mapping package to enable real-time communication.
    }
    \label{fig:taskaff}
\end{figure}

\subsection{Slicer-IPP}

The Slicer-IPP is composed of a data communication module, a prompt labeling module, and a visualization module. 
The data communication module accepts any volumetric image format, and it can pack them as image files used by SAM. 
The Slicer-IPP and SAM Server are designed to run five parallel tasks denoted as ``send inference request'' (SND\textunderscore{ }INF), ``receive inference request'' (RCV\textunderscore{ }INF), ``complete SAM inference'' (CPL\textunderscore{ }INF), ``receive mask transmission'' (RCV\textunderscore{ }MSK), and ``apply mask'' (APL\textunderscore{ }MSK). The affiliation of tasks is shown in Figure \ref{fig:taskaff}. The Slicer-IPP hosts SND\textunderscore{ }INF, RCV\textunderscore{ }MSK, and CPL\textunderscore{ }INF, while the server end hosts RCV\textunderscore{ }MSK and APL\textunderscore{ }MSK. Each task is executed synchronously as an independent loop.
All tasks run with the ``best-effort'' mode (opposite to ``guaranteed delivery'' mode), as the realtime-ness is the priority. Since 3D Slicer is a single-threaded software, each loop in the Slicer-IPP is set to have a 60 ms gap to process other tasks. A complete inference cycle starts from SND\textunderscore{ }INF and ends with APL\textunderscore{ }MSK. The time latency of one inference cycle is discussed in Section \ref{sec:result}.

\begin{figure}[h]
    \centering
    \includegraphics[width=0.85\textwidth]{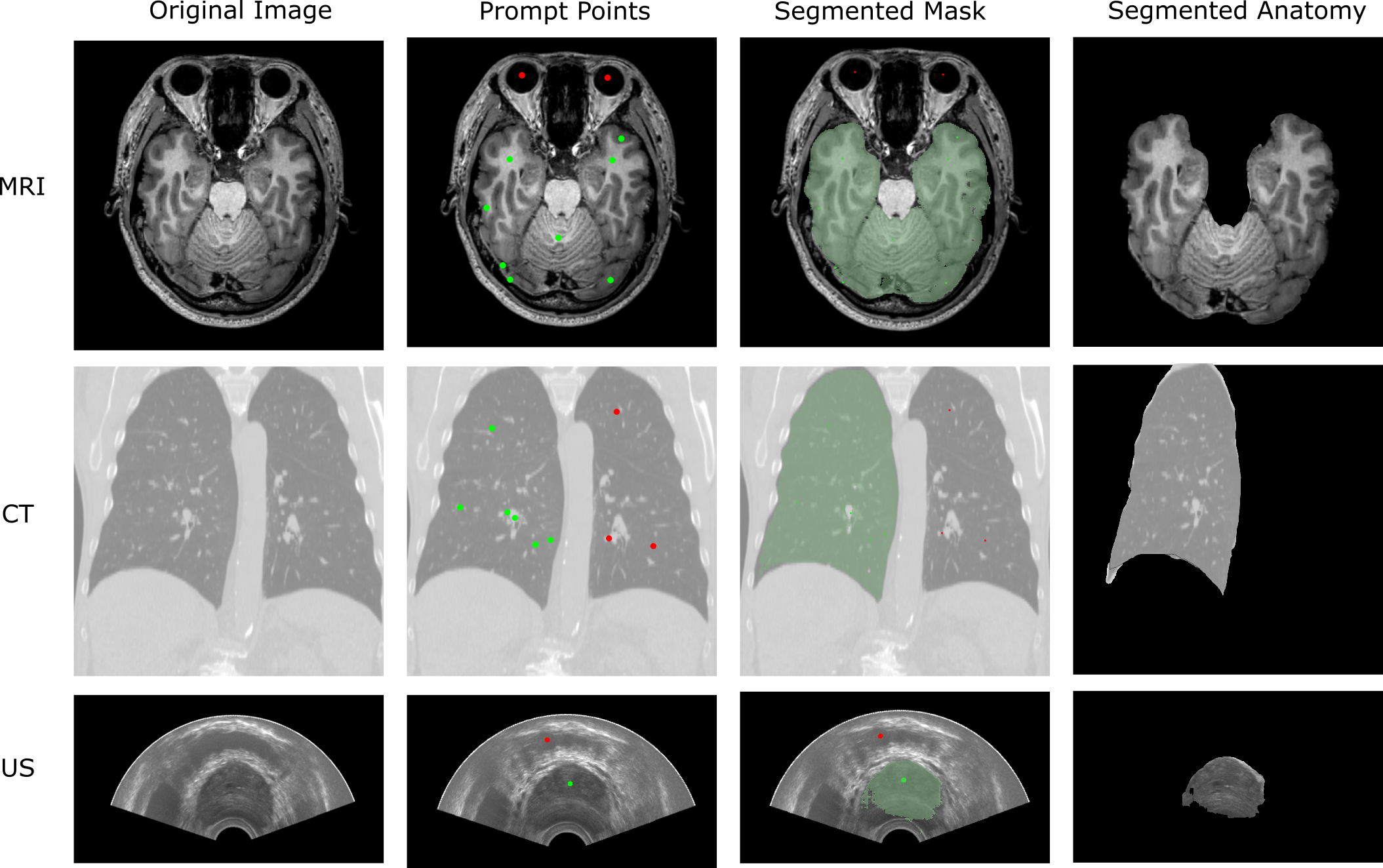}
    \caption{Example results for different image formats (CT, MRI, and US). The prompts with green points are for the regions to be selected, whereas the red points are for the regions to be removed. }
    \label{fig:diffmod}
\end{figure} 

To facilitate communication between the 3D Slicer and external tools or services, the platform uses ZeroMQ (ZMQ) \citep{2013zeromq} messaging library and Numpy \citep{numpy} memory mapping. 
ZMQ is a lightweight messaging library that enables high-performance, asynchronous communication between applications. 
In SAMM, ZMQ and Numpy are employed to transfer images, prompts, and requests between the Slicer-IPP and the SAM Server. The segmentation task is real-time by applying these two packages. 
This integration enables researchers to take advantage of SAM's cutting-edge segmentation capabilities within the familiar 3D Slicer platform, expanding the range of tasks that can be performed on images. 
The use of ZMQ and Numpy memory mapping also provides the flexibility to customize the communication protocol to fit the user's specific needs, further enhancing the versatility of the 3D Slicer platform. 

\begin{figure}[h]
    \centering
    \includegraphics[width=1.0\textwidth]{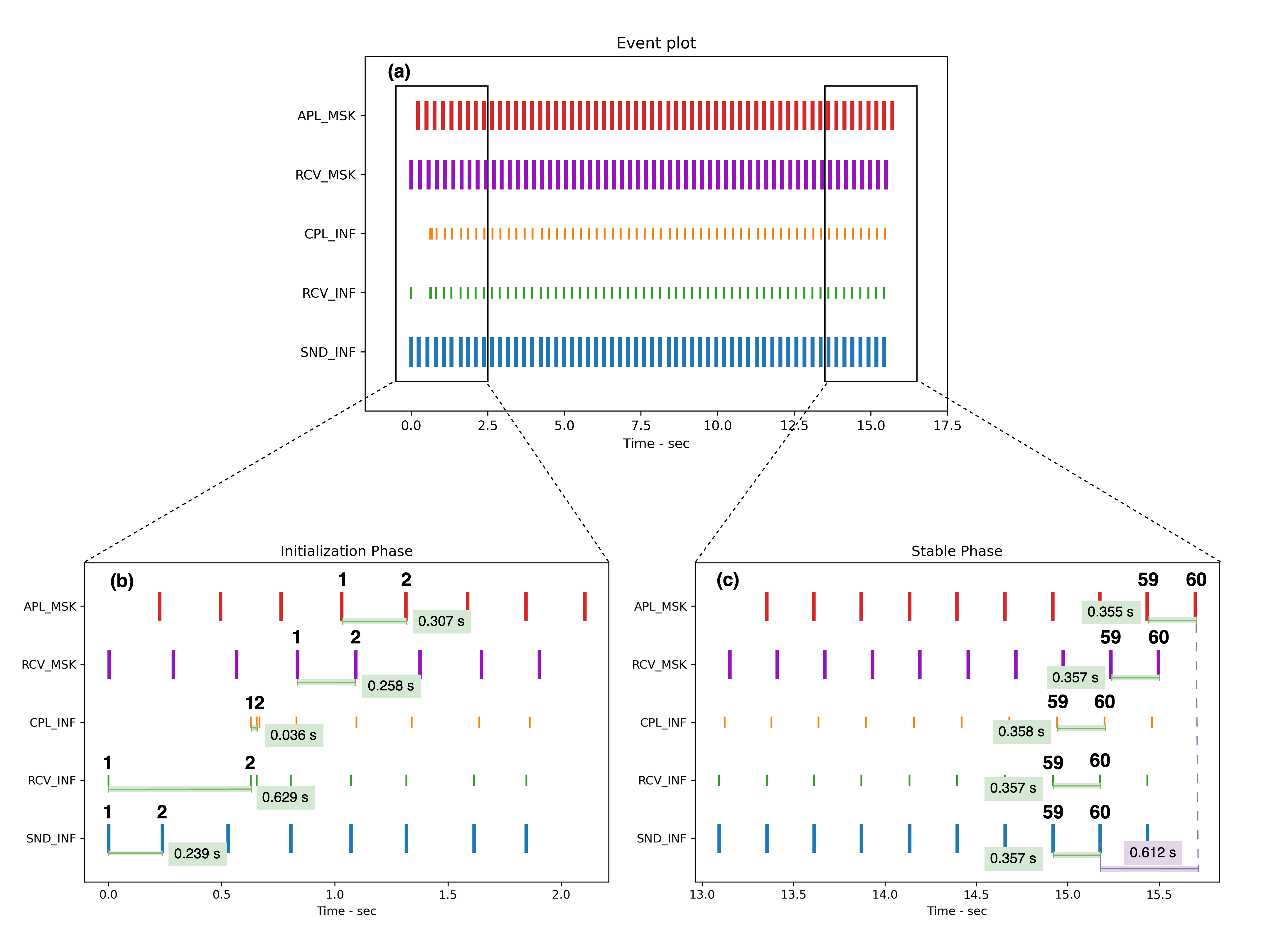}
    \caption{
    (a) is the event plot of the 5 tasks (SND\textunderscore{ }INF, RCV\textunderscore{ }INF, CPL\textunderscore{ }INF, RCV\textunderscore{ }MSK, and APL\textunderscore{ }MSK). Each row represents one task and each bar represents one event. A 16-second period (covers 60 cycles) from the initialization phase of SAMM (when all tasks are initially launched) to the stable phase (when all tasks run at a steady frequency) is shown. 
    Events executed within the same segmentation cycle are marked with the same number on the bottom two panels. The time interval of five tasks between event 1 and 2 is marked on (b) while the time interval of five tasks between event 59 and 60 is marked on (c). The time intervals between two cycles are shown in the green boxes whereas the time latency of one complete cycle is highlighted in the purple box.}
    \label{fig:synchronization}
\end{figure}

\section{Experiments and Results}\label{sec:result}

We evaluated the integration with SA-1B pre-trained model \footnote[1]{SA-1B is a model for SAM that is trained on 11 million diverse and high-resolution images and 1.1 billion high-quality segmentation masks \citep{kirillov2023segment}} for different formats including CT, MRI, and US, using images with different anatomies.
Figure \ref{fig:diffmod} shows the manually placed prompts and the segmentation masks generated from the sample datasets of 3D Slicer. 
The Slicer-IPP provides a built-in markup tool for placing the prompt points. The segmentation mask is overlaid on the original image once the CPL\textunderscore{ }INF task ends.
All cases are tested in the same environment (Ubuntu 20.04, AMD Ryzen 9 3900X, Nvidia GeForce RTX 3090).
The result demonstrates that SAM, although not specifically trained on medical image datasets, can generate masks for zero-shot segmentation tasks across different image formats. 

Here we use the term ``event'' to represent a task is completed. 
A complete cycle of the image segmentation consists of five events (Figure \ref{fig:synchronization}). 
For each time instance, the task owner logs the timestamp to a Python data storage object. The log data, generated in chronological order, is the output to a Python pickle file once 1000 segmentation cycles are completed. In Figure \ref{fig:synchronization}, only 60 complete cycles are shown. We evaluated the performance of the system using the end-to-end latency, defined as the time between a request of inference and the application of the inferred mask for the same image slice. In addition, the time intervals between every two consecutive tasks were measured.

In order to facilitate real-time interactive segmentation, the Slicer-IPP first computes the image embeddings and subsequently enables users to place prompt points. 
Therefore, the total execution time is divided into two components. The first component is the total time to compute the embeddings of all slices, and the second one is the latency per image in its runtime.

In the test environment, the embedding computation for a $352 \times 352 \times 240$ MRI image takes 162.9 seconds. 
For the same test dataset, the latency of an end-to-end process within the same segmentation cycle, including timestamp logging, is 0.612 seconds (measured at event 60 in the ``Stable Phase'', shown in Figure \ref{fig:synchronization}). 

\section{Discussion and Conclusion}

The integration of 3D Slicer and SAM enables researchers to conduct segmentation on medical images using a state-of-the-art foundation model. 
We have validated the capability of this model for segmenting images with a latency of 0.6 seconds. 
Our integration enables segmentation through prompts that can be automatically propagated to subsequent slices to streamline the segmentation process. 
This functionality could be particularly useful in large-scale image annotation when processing time is critical. 
 
In the future, the performance of SAM for segmentation of medical images can be enhanced by using medical image datasets and building upon the SA-1B generic model. With the SAM's support for text prompts \citep{kirillov2023segment}, future work may also include using text inputs as commands to perform segmentation tasks in 3D Slicer. Additionally, specialized medical AIs such as MONAI \citep{cardoso2022monai} may use SAM as an initialization tool to improve their inference and training efficiency.
 
As shown in Figure \ref{fig:synchronization}, we have noted that the initialization phase, especially at the SAM Server end, is not stable. The instability is evaluated by comparing the time latency of two consecutive events at the initialization and stable phase. The time latency of different tasks has a significant variation at the initialization phase but the instability typically disappeared in 5 seconds after initialization. Further work is needed to optimize the latency of the events in the segmentation cycles and the process of generating the image masks.

Overall, our results show that SAMM has a small latency when used as a prompt-based, semi-automatic segmentation method. 
The integration with 3D Slicer allows researchers to validate how SAM can be applied for segmentation of medical images. Moreover, the validation process can be significantly streamlined with the help of an open-source platform with numerous off-the-shelf tools.

\bibliographystyle{unsrtnat}
\bibliography{references}

\end{document}